\documentclass[superscriptaddress,prl,twocolumn,amsmath,twocolumn,amssymb]{revtex4-1}
\usepackage{graphicx}
\usepackage{color}
\usepackage{eurosym}
\usepackage{dcolumn}
\usepackage{bm}
\usepackage{subfigure}
\usepackage[final]{pdfpages}
\usepackage{natbib}
\graphicspath{{./images/}}

\newcommand{\ket}[1]{\left | #1 \right \rangle}

\begin{document}

\title{Heralding two- and four-photon path entanglement on chip}

\author{Jonathan C. F. Matthews}
\altaffiliation{These authors contributed equally}
\author{Alberto Politi}
\altaffiliation{These authors contributed equally}
\author{Damien Bonneau}
\author{Jeremy L. O'Brien}
\affiliation{Centre for Quantum Photonics, H. H. Wills Physics Laboratory \& Department of Electrical and Electronic Engineering, University of Bristol, Merchant Venturers Building, Woodland Road, Bristol, BS8 1UB, UK}
\email{Jeremy.OBrien@bristol.ac.uk}

\begin{abstract}
Generating quantum entanglement is not only an important scientific endeavor, but will be essential to realizing quantum-enhanced technologies, in particular, quantum-enhanced measurements with precision beyond classical limits. We investigate the heralded generation of multiphoton entanglement for quantum metrology using a reconfigurable integrated waveguide device in which projective measurement of auxiliary photons heralds the generation of path-entangled states. We use four and six-photon inputs, to analyze the heralding process of two- and four-photon NOON states-a superposition of N photons in two paths, capable of enabling phase supersensitive measurements at the Heisenberg limit. Realistic devices will include imperfections; as part of the heralded state preparation, we demonstrate phase superresolution within our chip with a state that is more robust to photon loss.
\end{abstract}

\maketitle

Quantum entanglement is understood to lie at the heart of proposed quantum technologies \cite{gi-nphot-1-165,la-nat-464-45,gi-sci-306-1330}. Entangling interactions between photons can be achieved using only linear optical circuits, additional photons and photon detection \cite{kn-nat-409-46,ob-sci-318-1567} where a particular detection event heralds the success of a given process. In this way it is possible to generate multi-photon entangled states and indeed to efficiently perform universal, fault tolerant quantum computing \cite{kn-nat-409-46}. There have been several examples of heralding two-photon \citep{ga-prl-93-020504,ok-sci-323-483,ba-prl-98-170502,zh-pra-77-062316,wa-natphot-4-549,*ba-natphot-4-553} and four-photon \cite{pr-prl-103-020503,*wi-prl-103-020504} polarization entanglement for quantum information processing applications. In the context of quantum metrology, however, generating path-number entangled states (including `NOON' states) is a particularly important example where an $N$-photon entangled state is heralded from $>N$ input photons and several schemes for doing this have been proposed \cite{le-pra-65-030101,ko-pra-65-052104,fi-pra-65-053818}. Here we use an integrated waveguide device to implement a scalable scheme for heralding path entangled states of up to four photons, including ones which are robust to losses. This scheme scales to arbitrary $N$ \cite{ca-prl-99-163604}.

Sub-wavelength sensitivity makes optical interferometry one of the most powerful precision measurement tools available to modern science and technology \cite{Mayinger}, with applications from microscopy to gravity wave detection \cite{ab-sci-256-325,go-nphys-4-472}. However, the use of classical states of light limits the phase precision $\Delta\theta$ of such measurements to the shot noise, or standard quantum limit (SQL): $\Delta\theta\cong1/\sqrt{N}$, where $N$ is the average number of sensing photons passing through the measurement apparatus. Quantum states of light---entangled states of photon number across the two paths of an interferometer for example---enable precision better than the SQL \cite{gi-sci-306-1330}. Quantum metrology promises to be of critical importance for applications where properties of the measured sample are altered by the sensing process: {by using entangled light, the same level of precision in measurement can be achieved by exposing the sample to fewer photons. Conversely, for the same disturbance of the sample (\emph{i.e.} the number of photons it is exposed to) more information can be extracted.

Entangled states of $M+N$ photons across two optical modes $x$ and $y$ of the form
\begin{eqnarray}
\ket{N::M}^{\alpha}_{x,y} = \tfrac{1}{\sqrt{2}}(\ket{N}_x\ket{M}_y + e^{i\alpha} \ket{M}_x\ket{N}_y)
\end{eqnarray}
can be used to increase the frequency of interference fringes by a factor of $\left|N-M\right|$ and to increase precision. The canonical example is the NOON state ($M=0$), which enables the ultimate precision $\Delta\theta \cong1/N$---the Heisenberg limit \cite{do-cphys-49-125}. While NOON states are fragile with respect to photon loss, other linear superpositions of photon number entanglement can beat the SQL in interferometers with loss:  states with $M\neq N$ are optimal for balanced loss \cite{do-pra-80-013825}. Realistic application of these entangled states, however, demands a scalable and practical means of generating large $\ket{N::M}$ states.

Multi-photon interference has been observed with post-selection of three-~\cite{mi-nat-429-161} four-~\cite{wa-nat-429-158,na-sci-316-726} and five-photon \cite{af-sci-328-879} states. To take advantage of the benefits of quantum metrology---whereby more information can be extracted for the same light intensity (photon flux through the sample) as a classical measurement---requires a scheme where the post-selection probability is sufficiently high \cite{ok-njp-10-073033}. To obtain the maximum precision, however, post-selection should be avoided, requiring either a deterministic or a heralded \cite{le-pra-65-030101,ko-pra-65-052104,fi-pra-65-053818,ca-prl-99-163604} method for generating high fidelity, large photon number $\ket{N::M}^{\alpha}_{x,y}$ states. In general, this requires auxiliary photons and photon detection \cite{kn-nat-409-46,ca-prl-99-163604}. A heralding scheme to generate NOON state of polarization entangled photons has been demonstrated for up to three photons \cite{ki-oe-17-19720}}. It is important to note that a heralding signal enables gating with an optical switch to expose the sample only to photons in the state $\ket{N::M}^{{\alpha}}_{x,y}$; the rate of production is given by the heralding probability but does not affect sensitivity.

For many precision measurement applications it is also important that the entangled state be encoded in two spatial modes (rather than polarization modes). Stability required for such encoding can be readily achieved in compact integrated quantum photonic devices \cite{po-sci-320-646}, as demonstrated by two-\cite{ma-natphot-3-346,sm-oe-17-13516} and four-photon \cite{ma-natphot-3-346} interference. While it is relatively straight forward to convert polarization entanglement to path entanglement in the bulk optical architecture, conversion is not so straightforward in the integrated optical architecture which, with the integration of photon sources and detectors, is of critical importance in bringing practical quantum technologies out of the research lab and into application.

\begin{figure}[t!]
\vspace{-0.25cm}
    \centering
    \includegraphics[width = 7.5cm]{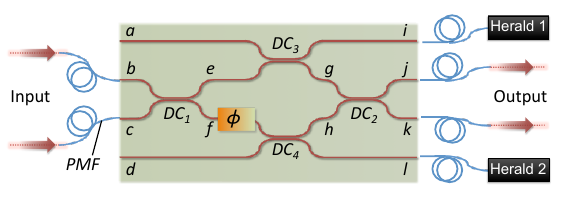}
    \vspace{-0.25cm}
\caption[]{\footnotesize{{Heralding multi-photon path-entangled states in a photonic chip.} The waveguide circuit  with coupling reflectivities $DC_{1,2}=1/2$, $DC_{3,4}=1/3$.}}
\label{fig_ExpSetup}
\vspace{-0.5cm}
\end{figure}

The silica-on-silicon waveguide circuit shown in Fig.~\ref{fig_ExpSetup}(a) is capable of heralding the two- and four-photon NOON states $\ket{2::0}_{j,k}^0$ and $\ket{4::0}_{j,k}^\pi$, as well as the four-photon state $\ket{3::1}_{j,k}^0$, dependent upon the input state and the setting of the internal phase $\phi$, as we now explain. The circuit consists of directional couplers $DC_{1-4}$, equivalent to beam splitters with lithographically defined reflectivity $\eta$, used to couple photons between optical modes and for quantum interference \cite{po-sci-320-646}. The resistive heating element controls the relative optical phase $\phi$ inside the device. The state $\ket{2::0}^0_{j,k}$ can be generated by inputting four frequency degenerate photons, via polarization maintaining fibre (PMF), in the (unentangled) path encoded state $\ket{2}_b\ket{2}_c$. Quantum interference at the first directional coupler $DC_1$---designed to have a reflectivity $\eta = 0.5$---generates a superposition of the components $\ket{4}_e\ket{0}_f$, $\ket{2}_e\ket{2}_f$  and $\ket{0}_e\ket{4}_f$. After $DC_3$ and $DC_4$ this state evolves to a superposition across the four modes $i$, $g$, $h$ and $l$. However, only the component $\ket{2}_e\ket{2}_f$ gives rise to terms that include $\ket{1}_i\ket{1}_l$. Detecting one and only one photon in each of these two heralding modes therefore projects the quantum state across modes $g$ and $h$ to $\ket{1}_g\ket{1}_h$. Quantum interference \cite{ho-prl-59-2044} at the final directional coupler $DC_2$ yields the two photon state $\ket{2::0}^0_{j,k}$. Provided $DC_3$ and $DC_4$ are $\eta = 0.5$, the intrinsic heralding success rate, \emph{i.e.} the probability of detecting $\ket{1}_i\ket{1}_l$ and thereby heralding $\ket{2::0}^0_{j,k}$, is $1/16$ (Ref. \cite{le-pra-65-030101}, see Supplemental Material).

For a low loss regime and in the absence of higher photon number terms, the heralding of the $\ket{1}_i\ket{1}_l$ component eliminates the lower order input state $\ket{1}_b\ket{1}_c$. We note that the requirements of heralding states for quantum metrology are more relaxed than for quantum computation or cryptography. A false heralded event of the vacuum state (due for example to a lower order input state $\ket{1}_b\ket{1}_c$) would be detrimental for any computation. In contrast, when low photon flux is the main requirement (exposure of a measured sample to radiation is to be kept to a minimum), a false heralding event of a vacuum state will not expose the sample to radiation.

The four-photon states $\ket{3::1}^{2\phi}_{g,h}$ and $\ket{4::0}^{\pi}_{j,k}$ are heralded in a similar manner: On inputting the state $\ket{3}_b\ket{3}_c$ of six frequency degenerate photons into the chip, non-classical interference at $DC_1$ yields a coherent superposition of the components $\ket{6}_e\ket{0}_f$, $\ket{4}_e\ket{2}_f$, $\ket{2}_e\ket{4}_f$  and $\ket{0}_e\ket{6}_f$. On detecting one photon in each of the two modes $i$ and $l$ (again via $DC_3$ and $DC_4$) projects the state into a superposition state $\ket{3::1}^{2\phi}_{g,h}$. With the phase set to $\phi = 0$, the state returns to $\ket{3::1}^{0}_{j,k}$ after quantum interference at $DC_2$. With the phase set to $\phi=\pi/2$, however, quantum interference at $DC_2$ yields the four photon NOON state  $\ket{4::0}^{\pi}_{j,k}$. For $\eta=0.5$ for both $DC_3$ and $DC_4$, the success rate of heralding $\ket{4::0}^{\pi}_{j,k}$ at the output is $3/64$ (Ref. \cite{le-pra-65-030101}, see Supplemental Material). Detection of the state $\ket{4}_j\ket{0}_k$, for example, leads to an interference fringe as a function of $\phi$, with resolution double that of classical light, providing an important means of testing the required quantum coherence within the optical circuit with respect to the heralding scheme.

Four- and six-photon input states were generated using a bulk optical 785nm, type-I pulsed down conversion source (see Supplemental Material) and coupled into the chip using polarisation maintaining fibre. Detection of multiple photon states at the output of the chip in the same optical mode is accomplished non-deterministically using cascaded non-number resolving, optical fibre-coupled single photon counting modules (SPCM, see Supplemental Material). The heralding process is tested with the assumption of conservation of photon number at all of the outputs of the device coupled to the SPCM detection scheme; using number resolving detectors at outputs $i$ and $l$, together with a deterministic photon source, would herald the generated entanglement without the need for counting all photons at the output of the device.

The phase instability of states leaving outputs $j$ and $k$ into a fibre or bulk optical circuit prevents a standard tomographic approach to reconstruct the density matrix of the state outside the chip. Development for an integrated optical quantum metrology device will incorporate the heralding circuit in one monolithic chip with all necessary components, including forming a waveguide interferometer for measuring unknown phase \cite{ma-natphot-3-346}. To analyse the isolated circuit, we have employed a series of steps to test the coherence and measure the relative photon number of the output state of the chip: (i) Temporal coherence of the multi-photon input states were verified with a generalised Hong-Ou-Mandel experiment to observe quantum interference of the state $\ket{2}_b\ket{2}_c$ incident on a 50\% reflectivity beamsplitter---we observed $V=34\pm4\%$ visibility interference in detecting two photons in each output of the beamsplitter which, compared to the ideal $V=1/3$ visibility, indicates temporal coherence of the multi-photon states generated in the photon source \cite{na-sci-316-726}; (ii) Photon number statistics were measured at the outputs $j$ and $k$ (equivalent to the diagonal of the density matrix) using non-deterministic number resolving detection with optical fibre splitters; (iii) The output state was then interfered on a second beamsplitter using a directional coupler inside the chip, via an inherently phase-stable fibre Sagnac loop.

\begin{figure}[tp!]
\vspace{-0.25cm}
    \centering
    \includegraphics[width = 8cm]{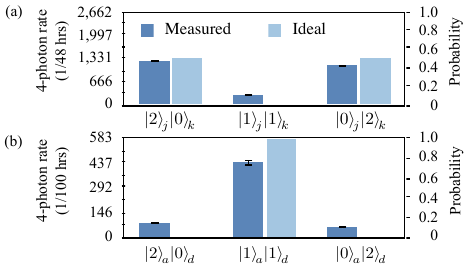}
    \vspace{-0.25cm}
\caption[]{\footnotesize{{Heralded $\ket{20}+\ket{02}$ state.} (a) Measurement of photon statistics of the heralded two-photon NOON state. (b) Testing coherence of the heralded two-photon NOON state by measuring the photon statistics after non-classical interference at $DC_2$ via a fibre Sagnac loop. Both distributions are normalised using single photon detection rates to account for relative detector scheme, source and waveguide coupling efficiencies. Error bars are given the standard deviation of detected events, assuming Poissonian statistics.}}
\vspace{-0.5cm}
\label{fig_2noondata}
\end{figure}

The photon number statistics measured from the heralded two-photon NOON state at outputs $j$ and $k$ is plotted in Fig.~\ref{fig_2noondata}(a). Probability-theoretic fidelity ($F = \sum_j \sqrt{p^e_j p^m_j}$) between the measured probability distribution
($p^m$) of photon statistics and the expected distribution ($p^e$) for the ideal state $\ket{2::0}^0_{j,k}$ (also plotted) is $F_i=0.95\pm0.01$.  For $\lambda =785\textrm{nm}$ operation, the reflectivities of $DC_1$ and $DC_2$ are measured to be $\eta=0.542$ and $0.530$ respectively. Using these measured reflectivities, and assuming otherwise perfect quantum interference, the expected output state was simulated; the fidelity between the photon number distribution of this simulated state and  the experimental results is $F_s=0.96\pm0.01$, leaving the discrepancy with perfect fidelity attributed to six- and higher photon number terms from the down conversion process and residual distinguishability of photons and not the device itself.

To test the coherence of the output of the circuit we formed a Sagnac loop by joining two optical fibres coupled to modes $j$ and $k$ (see Supplemental Material). This configuration results in quantum interference at $DC_2$ in the reverse direction and is equivalent to interference at a separate beamsplitter with zero relative optical phase of the two paths, fixed by the inherently stable Sagnac interferometer. By coupling detectors to waveguides $a$ and $d$, the photon statistics of the quantum state returning through the chip after $DC_2$ at $g$ and $h$ can be measured, with an intrinsic loss due to $DC_{3,4}$.
The fidelity between the measured distribution of photon statistics (Fig.~\ref{fig_2noondata}(b)) and the one expected from a perfect  $\ket{2::0}^0_{j,k}$ state interfering at directional coupler $DC_2$ is $F_i=0.90\pm0.03$. (Taking into account only the measured reflectivities of $DC_1$ and $DC_2$ the simulated detection rates agree with the experimental measurements with fidelity $F_s=0.97\pm0.03$.) Together with the temporal coherence of the input and the high fidelity of the output state in the diagonal basis, this demonstrates coherence of the $\ket{2::0}_{e,f}$ state.
\begin{figure}[tp!]
\vspace{-0.25cm}
    \centering
		 \includegraphics[width = 7.5cm]{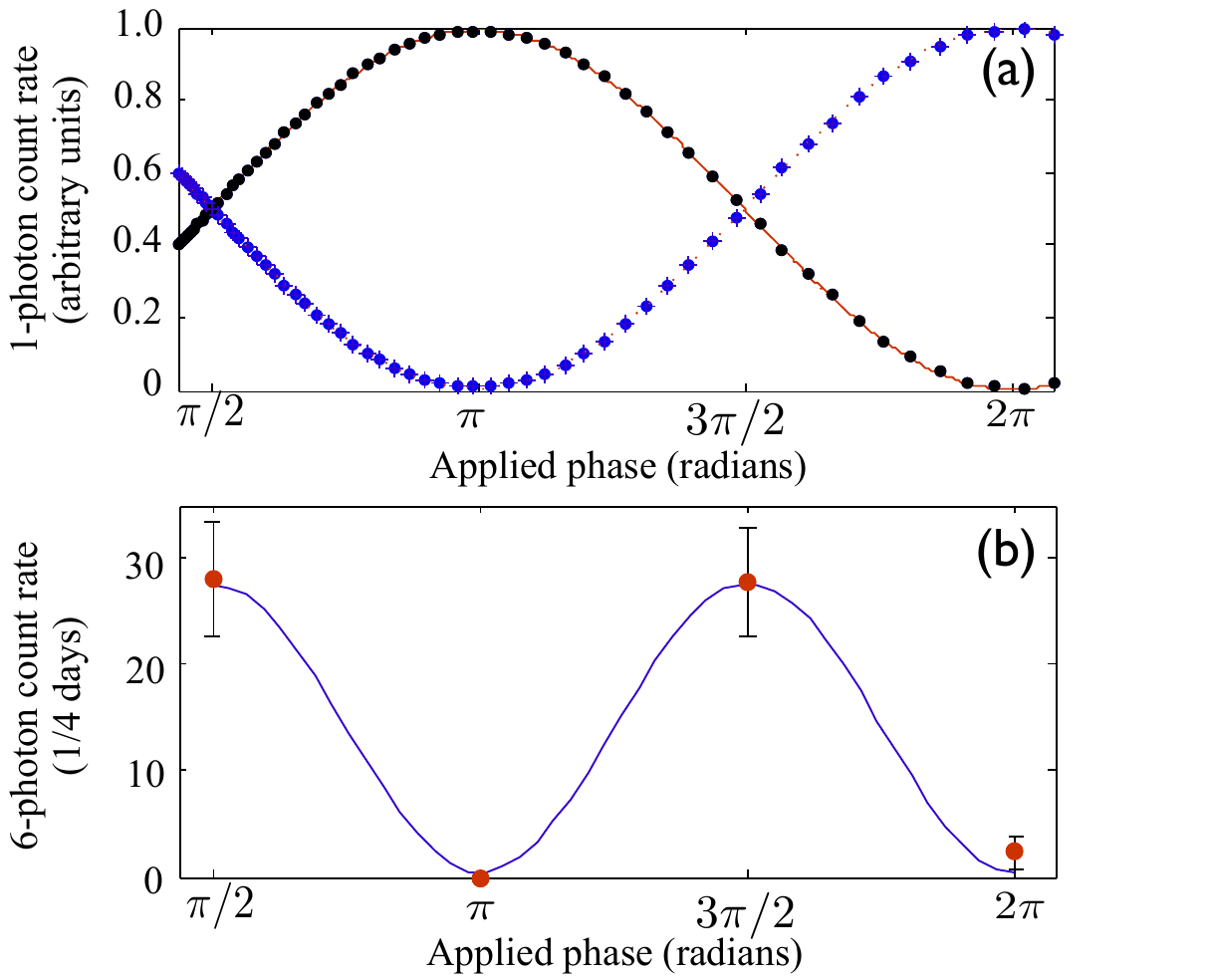}
		 \vspace{-0.25cm}
\caption[]{\footnotesize{{Super resolution with a heralded four photon entangled state.} (a) Single photon fringes from inputting light into waveguide $b$ and varying the phase $\phi$, displaying the expected pattern arising from classical interference pattern with period $2\pi$. Black circles and blue diamonds respectively represent the normalised single photons count rate detected at output $j$ and $k$. (b) The increased resolution interference pattern of manipulating $\phi$ of the state $\ket{3::1}_{e,f}^\phi$ with period $\pi$. The four data points represent six-photon count rates integrated over four days and normalized using single-photon count rates to account for coupling efficiency over time. Coincidence rates arising from higher photon number terms or otherwise were not subtracted from the data. Error bars are given the standard deviation of detected events, assuming Poissonian statistics. Blue sinusoidal plot of near unit contrast is plotted as a guide.}}
\vspace{-0.5cm}
\label{WedgeFringe}
\end{figure}

Although Fig.~\ref{fig_2noondata}(b) demonstrates coherence of the output state, the four photon process that generates it does not rely on phase stability within the interferometer structure of the device. In contrast heralding the $\ket{4::0}^0_{j,k}$ state from the six photon input state $\ket{3}_b\ket{3}_c$ requires coherent generation of the state  $\ket{3::1}^0_{g,h}$ within the interferometer. To test this coherence we injected the state $\ket{3}_b\ket{3}_c$ into the chip and sequentially set the phase to the four values $\phi  = \pi/2, \pi, 3\pi/2, 2\pi$. On detection of the six-photon state $\ket{1}_i\ket{4}_j\ket{0}_k\ket{1}_l$, we observed the sampled interference pattern plotted in Fig.~\ref{WedgeFringe}(b) which demonstrates two-fold super-resolution compared to the single photon interference pattern plotted in Fig.~\ref{WedgeFringe}(a), and coherence of the state $\ket{3::1}^0_{g,h}$ for subsequent generation of the $\ket{4::0}^0_{j,k}$ state. The small number of data points does not allow fitting to a sinusoidal fringe. Note that $\ket{3::1}^0_{g,h}$ is maximally entangled and is reported to be more robust to balanced loss, than the four-photon NOON state \cite{do-pra-80-013825}.

\begin{figure}[tp!]
\vspace{-0.25cm}
    \centering
    \includegraphics[width = 8cm]{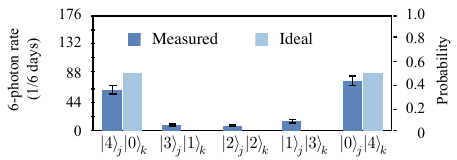}
    \vspace{-0.25cm}
\caption[]{\footnotesize{{Heralded generation of the $\ket{40}+\ket{04}$ state.}The distribution of photon statistics from measuring a heralded four-photon NOON state. The six-fold detection rates are normalized by single photon detection rates to account for relative source, coupling and detection scheme efficiencies. Error bars are given the standard deviation of detected events, assuming Poissonian statistics.}}
\vspace{-0.5cm}
\label{4NOON_data}
\end{figure}

Fig.~\ref{4NOON_data} shows the photon statistics of the $\ket{4::0}^{0}_{j,k}$ state that results from the quantum interference of the state $\ket{3::1}^{\pi/2}_{g,h}$ at $DC_2$. We fixed the phase within the chip to $\phi=\pi/2$ and again injected the six-photon state $\ket{3}_b\ket{3}_c$ into the chip. Six photons were detected in all possible four-photon combinations on outputs $j$ and $k$, together with a single photon in each of the heralding modes $i$ and $l$. The fidelity between the resulting distribution of photon statistics and the distribution expected from measuring the ideal state $\ket{4::0}^{0}_{j,k}$ is $F_i=0.89\pm0.04$. Taking into account the measured reflectivities of $DC_1$ and $DC_2$, the simulated statistics agree with experimental measurements with a fidelity $F_s=0.93\pm0.04$. The remaining discrepancy is attributed to a degree of distinguishability of the input photons---leading to imperfect quantum interference---and eight- and higher-photon number states from the pulsed down-conversion process that lead to different terms in the output state, allowing false heralded events: The eight-photon input $\ket{4}_b\ket{4}_c$ can give rise to the term $\frac{-i}{27\sqrt{3}}\ket{2}_i\ket{3::2}^{\pi/2}_{j,k}\ket{1}_l$ in the output state which in our experiment would be interpreted as a heralded ``$\ket{2}_j\ket{2}_k$'' event (see Supplemental Material).

The heralded generation of path entanglement will be crucial to the practical application of quantum metrology; the schemes presented here are scalable to arbitrary large entangled states \cite{ko-pra-65-052104,ca-prl-99-163604}. States that are robust to loss will be particularly important. The integrated waveguide architecture delivers the high stability and compact implementation required for real world applications. In particular, integrated variable beam splitters \cite{ma-natphot-3-346} will allow optimisation of quantum state engineering in the presence of loss \cite{do-pra-80-013825, ka-natphot-4-357}. The ongoing development of efficient number resolving detectors and deterministic photon sources such as single emitters or multiplexed down-conversion schemes\cite{ob-nphot-3-687}, shows promise for practical quantum metrology and other photonic quantum technologies when combined with circuits such as that described here. Real time quantum metrology requires high repetition rate (bright) sources of many photons. Future development will also require integration of fast feed-forward---using for example electro-optic materials---with the circuit demonstrated here to permit only intended quantum metrology states to interact with measured samples. This will likely form a building block for scalable generation of arbitrarily large entangled states \cite{ca-prl-99-163604,ca-arxiv-1103.2304}.
\vspace{0.5cm}
\begin{acknowledgments}
\footnotesize{We thank, J. P. Hadden A. Laing, A. Lynch, G. J. Pryde, J. G. Rarity, F. Sciarrino, A. Stefanov and X. Q. Zhou for helpful discussion. This work was supported by EPSRC, ERC, IARPA, the Leverhulme Trust, PHORPBITECH, QAP, Q-ESSENCE, QIP IRC, QUANTIP and NSQI. J.L.O'B. acknowledges a Royal Society Wolfson Merit Award.}
\end{acknowledgments}

\section*{APPENDIX}

\noindent\textbf{Detection scheme}

\begin{figure*}[t]
    \centering
    \includegraphics[width =16cm]{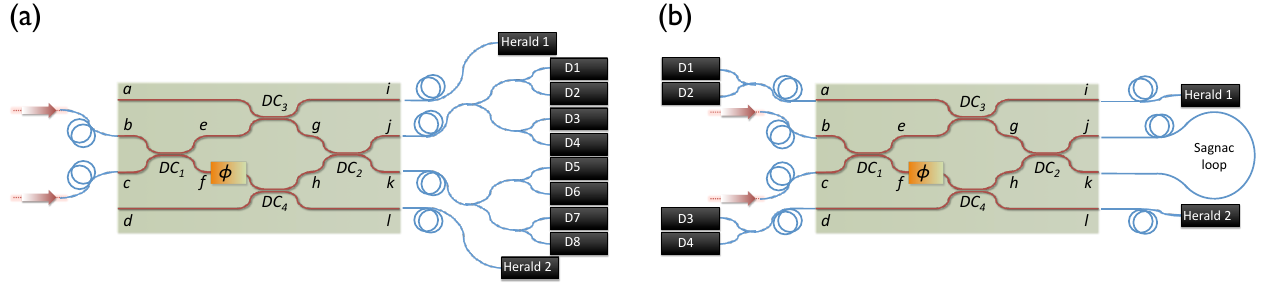}
\caption{\footnotesize{{Experimental setup for multi-photon state detection.}
(a) The detection scheme for detecting the six-photon state $\ket{1}_i\ket{4::0}^{\pi/2}_{j,k}\ket{1}_l$ at the output of the integrated chip. (b) The detection scheme for testing coherence of the nominally two-photon NOON state $\ket{1}_i\ket{2::0}_{j,k}\ket{1}_l$.}}
\label{fig_ExpSetup2}
\end{figure*}

Here we show in detail the schemes used to detect multi-photon states at the output of the integrated chip. The detectors used are silicon avalanche single photon counting modules (SPCM) that do not discriminate photon number---the presence of one of more photons at the SPCM produces the same output electrical signal. To reconstruct multi-photon states, number resolution is needed, which can be obtained probabilistically when using multiple SPCM and optical splitters. Three splitters and four detectors were used to detect up to four photons in the same optical mode as shown in the scheme in Fig.~S~\ref{fig_ExpSetup2}(a). The splitters used are multi-mode fiber couplers, with a near-unity transmissivity and close to 50:50 splitting ratio. If we assume perfect detectors and 50:50 splitters, the probability of detecting four photons in the same optical mode with the above method is given by $1/4^4 \times 4! = 3/32$. Similarly, the probability of detecting three photons in one mode and one in the other is $1/4^4 \times 3! = 3/128$, and the probability of detecting two photons in two modes is $1/4^4 \times 2! = 3/256$. All multi-photon coincidental detection reported in the Letter are normalized to the appropriate detection probability using single counts recorded in each detector; this normalizes the multi-photon coincidental detections, taking into account deviations from perfect and uniform detectors, differences in transmissivity and splitting ratio of the couplers.

Fig.~S~\ref{fig_ExpSetup2}(b) illustrates the detection scheme used to test coherence of the nominally $\ket{2::0}_{j,k}$ state. The measurement of the state at the end of the chip featured in Fig.~2 of the main text proves only that the state is mainly composed mainly by the components $\ket{20}_{j,k}$ and $\ket{02}_{j,k}$ but gives no information about the purity of the state. The coherence of the output state can be confirmed by interfering the photons in the paths $j$ and $k$ at a further directional coupler, since the result of this action is different in the case of a pure or a mixed state. This further interference cannot be achieved outside the integrated chip, otherwise the phase stability required for the experiment would be lost. We obtain non-classical interference at directional coupler $DC_2$, after the state is coupled out and back in the integrated chip via a fibre Sagnac loop between waveguides $j$ and $k$, ensuring complete phase stability of the photonic state; any variation in path registered by photons traveling from waveguide $j$ to $k$ is experienced also by photons traveling from waveguide $k$ to $j$. The photons are then probabilistically extracted from the chip via the directional couplers $DC_3$ and $DC_4$ and detected with cascaded detectors as for the other measurements.

\noindent\textbf{Photon coincidence counting}

\begin{figure}[t!]
\includegraphics[width =8cm]{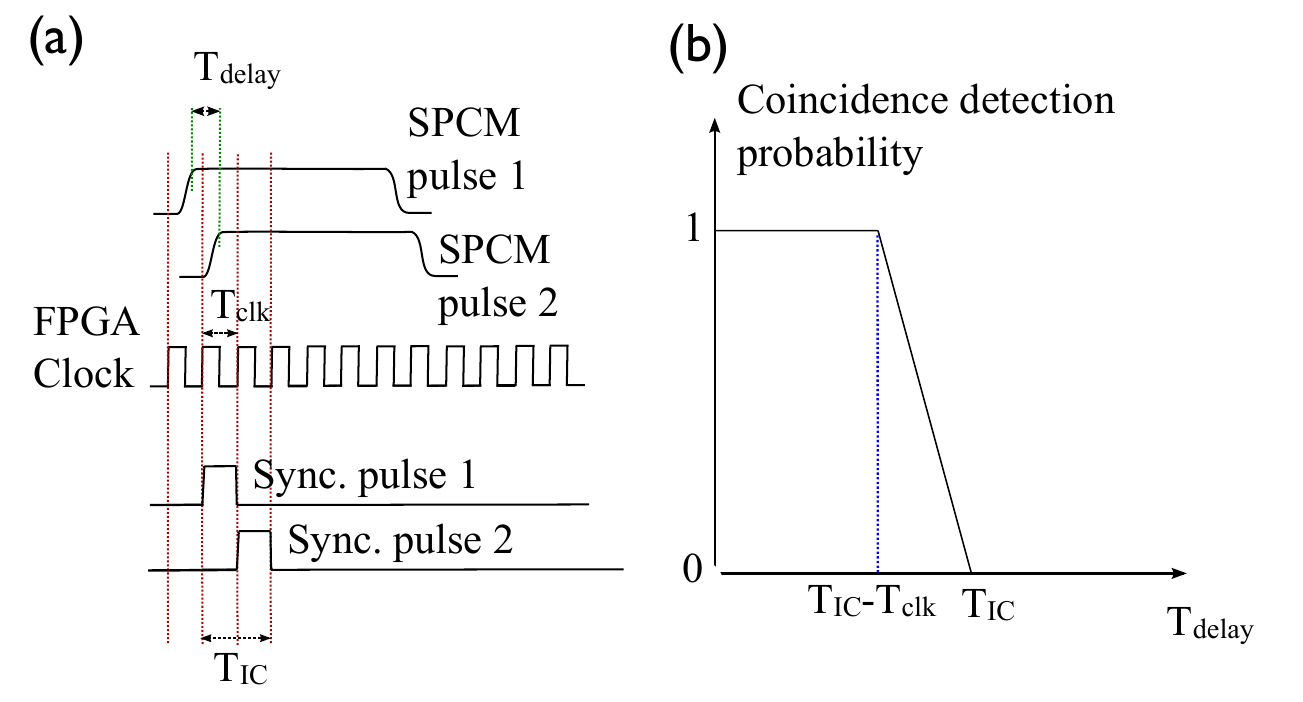}
\caption{\footnotesize{(a) An example of detection of coincidence for two channels 1 and 2. From the pulses of the detector SPCM pulse 1 (respspectively SPCM pulse 2), the synchronised signals Sync. pulse 1 (Sync pulse 2) are generated. A coincidence is then detected when two signals fall within $T_{IC}$. (b) The probability to detect a coincidence as a function of the delay between the two pulses.}}
\label{fig:Figure-coincidence}
\end{figure}

The multi-photon detection coincidences were registered and elaborated with a in-house FPGA virtex-4 board electronic circuit. The circuit works in the following way as illustrated on Fig.~S~\ref{fig:Figure-coincidence}(a). When a channel receives a pulse from the detector, the rising edge of the pulse is converted in an internal pulse synchronised with the FPGA clock. Such signal is then used to detect coincidences. We define an internal coincidence window $T_{IC}$ which is a multiple of the clock cycle $T_{Clk}$. If two or more synchronised signals from different channels fall within the window, then a coincidence is recorded for those events. Since the FPGA clock is not synchronised with the detector, the effective coincidence window does not have a rectangular shape as shown on Fig.~S~\ref{fig:Figure-coincidence}(b). In fact the probability to detect a coincidence is 100\% when the delay $T_{Delay}$ between the first and the last pulse is bellow $T_{IC}-T_{Clk}$. Then the probability to detect a coincidence falls linearly as $T_{Delay}$ increases from $T_{IC}-T_{Clk}$ to $T_{IC}$, with no coincidence detected when the delay is increased further. For this experiment, to account for the overall jitter of the six SPCM, we chose $T_{IC}=3T_{Clk}$ while the clock cycle of the FPGA is $T_{Clk}\approx2.9ns$. The counting logic has been checked using a Quantum Composer Plus 5218. We measured an 100\% efficiency detection window of 6ns and no detection after a delay of 9ns.

\noindent\textbf{Input state evolution}

The integrated circuit described in our Letter creates different quantum states depending on the input and on the value of the variable internal phase. Here we describe the evolution of the state for the inputs $\ket{2}_b\ket{2}_c$ and $\ket{3}_b\ket{3}_c$.
Each directional coupler $DC_i$ of reflectivity $\eta_i$ acting on two optical paths is modelled with the matrix
\begin{eqnarray}
DC_i \doteq \left(\begin{array}{c c} \sqrt{\eta_i} & i \sqrt{1-\eta_i} \\
										i \sqrt{1-\eta_i} & \sqrt{\eta_i} \end{array}\right)
\end{eqnarray}
with the assumption that coupler reflectivity in the circuit (Fig.~1(a) of the main text) have the values $\eta_1=\eta_2=1/2$ and $\eta_3=\eta_4=1/3$. The four waveguide input, four waveguide output chip featured in Fig.~1(a) of the main text is therefore modelled with the $4\times 4$ matrix product
\begin{widetext}
{\footnotesize{
\begin{eqnarray}
\left(
\begin{array}{cccc}
 1 & 0 & 0 & 0 \\
 0 & \sqrt{\eta_2} & i \sqrt{1-\eta_2} & 0 \\
 0 & i \sqrt{1-\eta_2} & \sqrt{\eta_2} & 0 \\
 0 & 0 & 0 & 1
\end{array}
\right)
\left(
\begin{array}{cccc}
 \sqrt{\eta_3} & i \sqrt{1-\eta_3} & 0 & 0 \\
 i \sqrt{1-\eta_3} & \sqrt{\eta_3} & 0 & 0 \\
 0 & 0 & \sqrt{\eta_4} & i \sqrt{1-\eta_4} \\
 0 & 0 & i \sqrt{1-\eta_4} & \sqrt{\eta_4}
\end{array}
\right)
\left(
\begin{array}{cccc}
 1 & 0 & 0 & 0 \\
 0 & 1 & 0 & 0 \\
 0 & 0 & e^{i \phi } & 0 \\
 0 & 0 & 0 & 1
\end{array}
\right)
\left(
\begin{array}{cccc}
 1 & 0 & 0 & 0 \\
 0 & \sqrt{\eta_1} & i \sqrt{1-\eta_1} & 0 \\
 0 & i \sqrt{1-\eta_1} & \sqrt{\eta_1} & 0 \\
 0 & 0 & 0 & 1
\end{array}
\right)
\end{eqnarray}
}}
\end{widetext}

When the state $\ket{2}_b\ket{2}_c$ is launched into the device, the directional coupler $DC_1$ transforms the state to
\begin{eqnarray}
\ket{2}_b\ket{2}_c  \stackrel{DC_1}{\rightarrow} - \sqrt{\frac{3}{4}}\ket{4::0}^{0}_{e,f} - \frac{1}{\sqrt{4}}\ket{2}_e\ket{2}_f.
\end{eqnarray}
The couplers $DC_3$ and $DC_4$, combined with the subsequent detection of only one photon each in waveguides $i$ and $l$, then project the state to
\begin{eqnarray}
\sqrt{\frac{4}{81}} e^{2 i \phi} \ket{1}_i\ket{1}_g\ket{1}_h\ket{1}_l.
\end{eqnarray}
The value of the variable phase $\phi$ is uninfluential in this case, since it can be treated as a global phase. Finally, this state non-classically interfers at directional coupler $DC_2$ to give the output
\begin{eqnarray}
\sqrt{\frac{4}{81}} e^{2 i \phi} \ket{1}_i\ket{2::0}^{0}_{j,k}\ket{1}_l,
\end{eqnarray}
corresponding to the two-photon NOON state $\ket{2::0}^0_{j,k}$ at the output of the photonic chip with the global phase $2\phi$.

Similarly, when the state $\ket{3}_b\ket{3}_c$ is launched into the photonic circuit, non-classical interference at $DC_1$ transforms the state according to
\begin{eqnarray}
\ket{3}_b\ket{3}_c & \stackrel{DC_1}{\rightarrow} & -\sqrt{\frac{5}{8}}\ket{6::0}^{0}_{e,f} - \sqrt{\frac{3}{8}}\ket{4::2}^{0}_{e,f}.
\end{eqnarray}
The couplers $DC_3$ and $DC_4$, combined with the subsequent detection of only one photon each in waveguides $i$ and $l$, project the state to
\begin{eqnarray}
i e^{i 2\phi}\sqrt{\frac{4}{243}}\ket{1}_i\left(\frac{\ket{3}_g\ket{1}_h+e^{2i\phi}\ket{1}_g\ket{3}_h}{\sqrt{2}}\right)\ket{1}_l.
\end{eqnarray}
The state $\ket{3::1}^{0}_{j,k}$ is created inside the photonic circuit after couplers $DC_3$ and $DC_4$. The variable internal phase $\phi$ (controllable via thermo-optical electrode) can be used to control this state, with the effect of non-classical interference at directional coupler $DC_2$ depending on the phase $\phi$ according to
\begin{eqnarray}
i e^{i 3 \phi} \sqrt{\frac{4}{243}}\ket{1}_i\left(\sin{\phi}\ket{4::0}^{\pi}_{j,k}-\cos{\phi}\ket{3::1}^{0}_{j,k}\right)\ket{1}_l.
\end{eqnarray}
It is noted that the internal phase $\phi$ is useful for two important tasks: confirming the quantum nature of the states inside the photonic chip, allowing the measurement of fringes with double the frequency of the one photon case reported in Fig.~3 of the main text; and for producing the desired state $\ket{4::0}_{j,k}$ at the output of the chip---as shown in Fig.~4 in the main text---with the choice $\phi=\pi/2$.
\\

\noindent\textbf{The multi-photon down conversion source}

\begin{figure}[htp!]
\includegraphics[width =8cm]{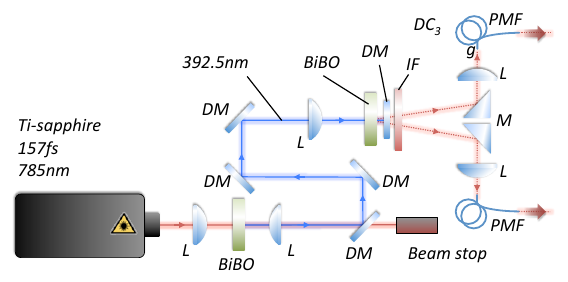}
\caption{\footnotesize{Schematic of the type-I, pulsed down conversion source used to generate four- and six-photon (unentangled) path encoded states.}}
\label{fig:source_schematic}
\end{figure}

Four- and six-photon input states were generated using the bulk optical type-I pulsed down conversion source given in Fig.~S\ref{fig:source_schematic}. Pulsed $\lambda=785$ nm light from a 157 fs, 80 MHz mode-locked Titanium:Sapphire laser system was up converted using a 2 mm nonlinear bismuth borate $\textrm{BiB}_3\textrm{O}_6$ (BiBO) crystal; the resulting $\lambda=392.5$ nm light is separated from remaining infrared light using four dichroic mirrors (DM) and focused (L) to seed type-I spontaneous parametric downconversion in a second BiBO crystal. The photons created in this process pass through a high transmission interference filter (IF) with FWHM = 2.5~nm and are collected by focusing two points on opposite sides of the down conversion cone onto PMFs which are butt-coupled to the waveguide chip. Beam waists of $\omega_0\sim40\mu\textrm{m}$ were chosen for the focused light of both up- and down-conversion BiBO crystals.

\noindent\textbf{Higher photon number contributions}

The photonic input needed for the 6-photon experiment to produce the states $\ket{3::1}^{0}_{j,k}$ and $\ket{4::0}^{\pi/2}_{j,k}$ corresponds to the state $\ket{3}_b\ket{3}_c$. Experimentally, an approximation of this state can be created via parametric down conversion---using the source given in Fig.~S\ref{fig:source_schematic}---and it can be written as
\begin{eqnarray}
\ket{\psi}_{bc} \sim \ket{00} + \xi \ket{11} + \xi^2 \ket{22} + \xi^3 \ket{33} + \xi^4 \ket{44} + ...\nonumber\\
\label{eq_SPDC}
\end{eqnarray}
As described in the text, the $\ket{00}$ and $\ket{11}$ states are rejected by detection of two heralding photons. For the six photon experiments other components with a number of total photons lower than six (i.e. $\ket{22}$) have no effect, since they cannot give rise to six-photon simultaneous events. On the other hand, the input component with eight photons can give a recordable event, since losses and detectors without photon number resolution wash out the information about the input state.

The effect of the state $\ket{4}_b\ket{4}_c$ acting as an input of the integrated chip is analyzed here. Non-classical interference at $DC_1$ transforms the state to
\begin{eqnarray}
\ket{4}_b\ket{4}_c\stackrel{DC_1}{\rightarrow}\frac{\sqrt{35}}{8}\ket{8::0}^{0}_{e,f}+\frac{\sqrt{5}}{4}\ket{6::2}^{0}_{e,f} + \frac{3}{8}\ket{4}_e\ket{4}_f\nonumber\\
\end{eqnarray}
To understand what happens if the eight-photon term is present in the circuit in the detector configuration adopted in this experiment, it should be noted that eight photons can give rise to six-photon coincidental detection in different ways. Since the general analysis is rather complex, we limit here to the case of $\phi = \pi/2$. For this phase, the complete state that can give rise to allowed events is of the form
\begin{eqnarray}
\frac{i\sqrt{2}}{162}\ket{1}_i\left(3\sqrt{5}\ket{6::0}^0_{j,k}-\sqrt{3}\ket{4::2}^0_{j,k}\right)\ket{1}_l&&\nonumber\\
+\frac{\sqrt{2}}{54\sqrt{3}}\ket{2}_i\left(-3\sqrt{5}\ket{5::0}^{-\pi/2}_{j,k}-i\ket{4::1}^{\pi/2}_{j,k}\right.&&\nonumber\\
\left.+\sqrt{2}\ket{3::2}^{-\pi/2}_{j,k}\right)\ket{1}_l&&\nonumber\\
+\frac{\sqrt{2}}{54\sqrt{3}}\ket{1}_i\left(3i\sqrt{5}\ket{5::0}^{\pi/2}_{j,k}+\ket{4::1}^{-\pi/2}_{j,k}\right.&&\nonumber\\
\left.-i\sqrt{2}\ket{3::2}^{\pi/2}_{j,k}\right)\ket{2}_l&&\nonumber\\
+\frac{2}{162}\ket{2}_i\left(-7\sqrt{3}\ket{4::0}^0_{j,k} +3\ket{2}_j\ket{2}_k\right)\ket{2}_l&&
\end{eqnarray}
It is clear that the eight-photon input term can give rise to a quite complex pattern of detection. In particular, all the terms apart from the first of each row can give rise to detection of the $\ket{3::1}_{j,k}$ and $\ket{22}_{j,k}$ state.

To minimize the effect of the higher order emission of the BiBO down-conversion crystal, the value of $\xi$ was chosen to be $\xi \sim 0.085$, that corresponds to a power of the blue beam pumping the down-conversion crystal of $P_b=215$~mW. This choice of $\xi$ is a good compromise between the six-photon detection rate and the unwanted production of eight photons. This is confirmed by Fig.~4 of the main text, since the count rates for the states $\ket{3::1}_{j,k}$ and $\ket{22}_{j,k}$ is low in comparison to the $\ket{4::0}_{j,k}$ term. To obtain higher quality states a true $\ket{3}_b\ket{3}_c$ could be be used. The generation of these Fock-states is in principle possible. Although being a complex problem, Fock states can be generated using different methods, for example linear optics \cite{sanaka} or atom-cavity systems \cite{brattle}.

%

\vspace{-0.75cm}

\vspace{-1cm}
\end{document}